\documentclass[aps,prd,nofootinbib,showpacs,twocolumn]{revtex4} %superscriptaddress
\usepackage{txfonts}
\usepackage{graphicx}
\usepackage{dcolumn}
\usepackage{bm}
\usepackage{amssymb}
\usepackage{latexsym}
\usepackage[colorlinks, linkcolor=blue, citecolor=blue, urlcolor=blue]{hyperref}

\newcommand{\be}{\begin{equation}}
\newcommand{\ee}{\end{equation}}
\newcommand{\bq}{\begin{eqnarray}}
\newcommand{\eq}{\end{eqnarray}}

\begin{document}
%\begin{CJK*}{GB}{} % Use default fonts from CJK (see below)
%\begin{CJK*}{GBK}{kai}

\title{Heal the world: Avoiding the cosmic doomsday in
the holographic dark energy model}

\author{Xin Zhang}
\email{zhangxin@mail.neu.edu.cn} \affiliation{Department of Physics,
College of Sciences, Northeastern University, Shenyang 110004,
China} \affiliation{Kavli Institute for Theoretical Physics China,
Chinese Academy of Sciences, Beijing 100080, China}

\begin{abstract}

The current observational data imply that the universe would end
with a cosmic doomsday in the holographic dark energy model.
However, unfortunately, the big-rip singularity will ruin the
theoretical foundation of the holographic dark energy scenario. To
rescue the holographic scenario of dark energy, we employ the
braneworld cosmology and incorporate the extra-dimension effects
into the holographic theory of dark energy. We find that such a mend
could erase the big-rip singularity and leads to a de Sitter finale
for the holographic cosmos. Therefore, in the holographic dark
energy model, the extra-dimension recipe could heal the world.

\end{abstract}


\maketitle
%\end{CJK*}

\section{Introduction}\label{sec:intro}

Dark energy was found by the observations of type Ia supernovae in
1998 \cite{Riess:1998cb}. It is believed that the current cosmic
acceleration is caused by dark energy. The basic characteristic of
dark energy is that its equation of state parameter $w$ (the
definition of $w$ is $w=p/\rho$, where $\rho$ is the energy density
and $p$ is the pressure) has a negative value ($w<-1/3$). At
present, the universe is under the domination of dark energy. The
combined analysis of cosmological observations shows that the
universe is spatially flat and consists of about $70\%$ dark energy,
$30\%$ dust matter (cold dark matter plus baryons), and negligible
radiation \cite{wmap5}.

Although it can be affirmed that the ultimate fate of the universe
is determined by the feature of dark energy, the nature of dark
energy as well as its cosmological origin still remain enigmatic at
present. The preferred candidate for dark energy is the famous
Einstein's cosmological constant $\lambda$ \cite{Einstein:1917ce},
however, it always suffers from the ``fine-tuning'' and ``cosmic
coincidence'' puzzles. The fine-tuning problem asks why the vacuum
energy density today is so small compared to typical particle
scales. The vacuum energy density is of order $10^{-47} {\rm
GeV}^4$, which appears to require the introduction of a new mass
scale 14 or so orders of magnitude smaller than the electroweak
scale. The second difficulty, the cosmic coincidence problem, asks:
Since the energy densities of vacuum energy and dark matter scale so
differently during the expansion history of the universe, why are
they nearly equal today? To get this coincidence, it appears that
their ratio must be set to a specific, infinitesimal value in the
very early universe. Lots of efforts have been made to try to
resolve the cosmological-constant problem, all these efforts,
however, turned out to be unsuccessful
\cite{Weinberg1989,Bousso:2007gp}.

On the other hand, at the phenomenological level, many dynamical
dark energy models have been proposed to interpret the observational
data (for reviews, see, e.g., Refs.~\cite{DErev}). In terms of the
equation of state, these dynamical dark energy scenarios can be
classified into the following three categories: quintessence,
phantom, and quintom. For quintessence-like dark energy, the
equation of state parameter is always greater than $-1$, namely
$w\geq -1$; for phantom-like dark energy, the equation of state
parameter is always less than $-1$, namely $w\leq -1$; for
quintom-like dark energy, the equation of state parameter crosses
$-1$ during the evolution. In the scenario of phantom dark energy,
it is remarkable that all the energy conditions in general
relativity (including the weak energy condition) are violated. A
ghastful prediction of this scenario is the ``cosmic doomsday''
\cite{bigrip}. Due to the equation of state less than $-1$, the
phantom component leads to a ``big rip'' singularity at a finite
future time, at which all bound objects will be torn apart. For
detailed discussions on the properties of future singularities of
the universe, see Ref.~\cite{Nojiri:2005sx}.

Possibly, the cosmological constant (or the vacuum energy density)
might also have some dynamical property. It is well known that the
cosmological constant is actually closely related to an ultraviolet
(UV) problem in the quantum field theory. A simple evaluation in
quantum field theory leads to a discrepancy of 120 orders of
magnitude between the theoretical result and the observational one
\cite{Weinberg1989}. Obviously, the key point is the gravity. In a
real universe, the effects of gravity should be involved in this
evaluation. So, actually, the cosmological constant (or dark energy)
problem is in essence an issue of quantum gravity
\cite{Witten:2000zk}. However, by far, we have no a complete theory
of quantum gravity, so it seems that we have to consider the effects
of gravity in some effective quantum field theory in which some
fundamental principles of quantum gravity should be taken into
account. It is commonly believed that the holographic principle
\cite{holop} is just a fundamental principle of quantum gravity.
Taking the holographic principle into account, dynamical vacuum
energy is possible.

The holographic principle is expected to play an important role in
dark energy research. When considering gravity, namely, in a quantum
gravity system, the conventional local quantum field theory will
break down due to the too many degrees of freedom that would cause
the formation of black hole. So, there is a proposal saying that the
holographic principle may put an energy bound on the vacuum energy
density, $\rho_{vac} L^3 \leq M^2_{pl}L$, where $\rho_{vac}$ is the
vacuum energy density and $M_{pl}$ is the reduced Planck mass
\cite{Cohen:1998zx}. This bound says that the total energy in a
spatial region with size $L$ should not exceed the mass of a black
hole with the same size. The largest size compatible with this bound
is the infrared (IR) cutoff size of this effective field theory.
Evidently, this bound implies a UV/IR duality. Therefore, the
holographic principle may lead to a dark energy model that is
actually based on the effective quantum field theory with a UV/IR
duality. From this UV/IR correspondence, the UV problem of dark
energy can be converted into an IR problem.

By phenomenologically introducing a dimensionless parameter $c$, one
can saturate that bound and write the dark energy density as
$\rho_{de}=3c^2M_{pl}L^{-2}$. The parameter $c$ is
phenomenologically introduced to characterize all of the
uncertainties of the theory. Now, the problem becomes how to choose
an appropriate IR cutoff for the theory. A natural choice is the
Hubble length of the universe, however, it has been proven that
there is no cosmic acceleration for this choice \cite{Hsu:2004ri}.
Li proposed that, instead of the Hubble horizon, one can choose the
event horizon of the universe as the IR cutoff of the theory
\cite{Li:2004rb}. This choice not only gives a reasonable value for
dark energy density, but also gives rise to an acceleration solution
for the cosmic expansion.

The parameter $c$ in the holographic dark energy model plays a very
important role in determining the final fate of the universe
\cite{holode1,holode2}. In particular, when $c$ is less than 1, the
equation of state of holographic dark energy will evolve across the
cosmological-constant boundary $w=-1$. Note that it will evolve from
the region of $w>-1$ to that of $w<-1$, so the choice of $c<1$ makes
the holographic dark energy finally become a phantom energy that
would lead to a cosmic doomsday (``big rip'') in the future. It
should be pointed out that the holographic dark energy model has
been strictly constrained by the current cosmological observations
\cite{obsholode1,obsholode2,holoobs09}. The joint analysis of the
latest observational data, including type Ia supernovae (SN), cosmic
microwave background (CMB), and baryon acoustic oscillation (BAO),
shows that the parameter $c$ is indeed less than 1 (at nearly 2
$\sigma$ level):
$c=0.818^{+0.113}_{-0.097}~(1\sigma)~^{+0.196}_{-0.154}~(2\sigma)$
\cite{holoobs09}. Thus, it seems that the big rip is inevitable in
the holographic dark energy model. However, on the other hand, in
the framework of holographic dark energy model, the big rip is
actually not allowed, due to the reason that the Planck scale
excursion of UV cutoff in the effective field theory is forbidden.
So, the occurrence of the cosmic doomsday would canker the
theoretical root of the holographic dark energy scenario.

To rescue the holographic dark energy model, we have to try to find
out some working mechanism to erase the big-rip singularity in the
phantom regime of the holographic dark energy scenario. In this
paper, we will explore such a mechanism remedied by which the
holographic theory of dark energy can be healed. We find that the
extra dimension mechanism would provide us with a good cure for the
illness of the holographic dark energy model. In what follows, we
will incorporate the extra dimension effects into the holographic
dark energy scenario, and we will see that such a mend works very
well: the big-rip singularity would be eliminated successfully.
Furthermore, we will find that the ultimate fate of the cosmos is an
attractor where the steady state (de Sitter) finale occurs.

This paper is organized as follows: In Sec.~\ref{sec:holode}, we
briefly review the holographic dark energy model and show that the
cosmic doomsday seems inevitable in this model according to the
current cosmological observations. We also expatiate on the
necessity of eliminating the big-rip singularity in the holographic
model of dark energy. In Sec.~\ref{sec:brane}, we employ the
braneworld mechanism to rescue the holographic dark energy model. We
show that the big-rip singularity would be erased successfully by
using the extra dimension mechanism, and the cosmic destiny would be
a de Sitter phase. Furthermore, we give possible constraint on the
holographic dark energy model from the extra dimension prescription.
Finally, we give the conclusion in Sec.~\ref{sec:conclusion}.

\section{Cosmic doomsday in holographic dark
energy scenario}\label{sec:holode}

The holographic dark energy model proposed by Li \cite{Li:2004rb} is
based on the future event horizon as an IR cutoff. The dark energy
density is written as
\begin{equation}
\rho_{de}=3c^2M_{pl}^2R_{eh}^{-2},\label{density}
\end{equation}
where $R_{eh}$ is the event horizon of the universe, which is
defined as
\begin{equation}
R_{eh}(t)=a(t)\int_t^\infty{dt'\over a(t')}.\label{ehorizon}
\end{equation}

From the definition of the event horizon (\ref{ehorizon}), we can
easily derive
\begin{equation}
\dot{R}_{eh}=HR_{eh}-1.\label{dotReh}
\end{equation}
So, taking derivative of Eq.~(\ref{density}) with respect to time
$t$ and using the energy conservation equation
$\dot{\rho}_{de}+3H(1+w)\rho_{de}=0$, we can obtain the equation of
state of holographic dark energy,
\begin{equation}
w=-{1\over 3}-{2\over 3c}\sqrt{\Omega_{de}},\label{eos}
\end{equation}
where
\begin{equation}
\Omega_{de}={\rho_{de}\over 3M_{pl}^2H^2}={c^2\over
H^2R_{eh}^2}\label{Ode}
\end{equation}
is the fractional density of holographic dark energy. For
convenience, hereafter, we will use the units with $M_{pl}=1$, but
we will still explicitly write out $M_{pl}$ at several places. To
see the evolution dynamics of the holographic dark energy, we take
derivative of Eq.~(\ref{Ode}) with respect to $\ln a$, and derive
\begin{equation}
\Omega_{de}'=2\Omega_{de}\left(\epsilon-1+{\sqrt{\Omega_{de}}\over
c}\right),\label{Odep}
\end{equation}
where $\epsilon\equiv -{\dot{H}/ H^2}=-{H'/ H}$, and a prime denotes
the derivative with respect to $\ln a$. Using the Friedmann equation
$3H^2=\rho_m+\rho_{de}$ and the equation of state of dark energy
(\ref{eos}), we have
\begin{equation}
\epsilon={3\over 2}(1+w_{de}\Omega_{de})={3\over
2}-{\Omega_{de}\over 2}-{\Omega_{de}^{3/2}\over c}.\label{epsilon}
\end{equation}
Hence, we obtain the equation of motion, a differential equation,
for $\Omega_{de}$,
\begin{equation}
\Omega _{de}^{\prime}=\Omega _{de}(1-\Omega
_{de})\left(1+\frac{2}{c}\sqrt{\Omega _{de}}\right).  \label{eom}
\end{equation}

\subsection{Cosmic doomsday}\label{subsec:bigrip}

The parameter $c$ plays a significant role for the cosmological
evolution of the holographic dark energy. When $c\geq 1$, the
equation of state of dark energy will evolve in the region of
$-1\leq w\leq -1/3$. In particular, if $c=1$ is chosen, the behavior
of the holographic dark energy will be more and more like a
cosmological constant with the expansion of the universe, such that
ultimately the universe will enter the de Sitter phase in the
distant future. When $c<1$, the holographic dark energy will exhibit
a quintomlike evolution behavior (for ``quintom'' dark energy, see,
e.g., Refs.~\cite{quintom} and references therein); i.e., the
equation of state of holographic dark energy will evolve across the
cosmological-constant boundary $w=-1$ (actually, it evolves from the
region with $w>-1$ to that with $w<-1$). That is to say, the choice
of $c<1$ makes the holographic dark energy behave as a quintom
energy that would lead to a cosmic doomsday (``big rip'') in the
future. Thus, as discussed above, the value of $c$ determines the
destiny of the universe in the holographic dark energy model.

%%%%%%%%%%%%%%%%%%%%%%%%%%%%%%%%%%%%%%%%%%%%%%%%%%%%%%%%%%%%%%%%%%
\begin{figure}[htbp]
\begin{center}
\includegraphics[scale=0.35]{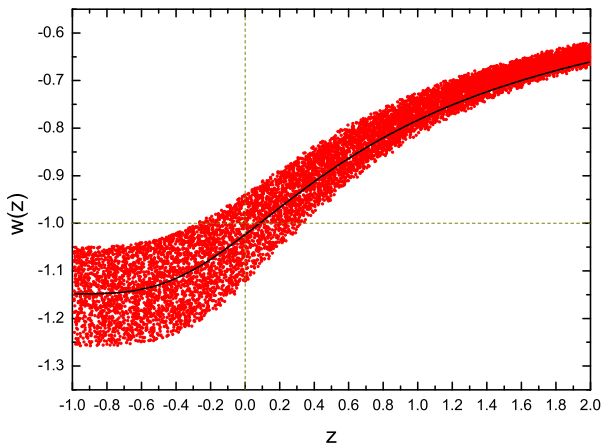}
\caption{The evolution of the equation of state of holographic dark
energy $w(z)$ with 1$\sigma$ uncertainty. In this figure, the
central black solid line represents the best fit, and the red dotted
area around the best fit covers the range of 1$\sigma$ errors. The
errors quoted here are calculated using a Monte Carlo method where
random points are chosen in the 1$\sigma$ region of the parameter
space: $c=0.818^{+0.113}_{-0.097}$ and
$\Omega_{m0}=0.277^{+0.022}_{-0.021}$. From this figure, one can
clearly see that our universe is striding into the phantom regime of
the holographic dark energy scenario.}\label{fig:wz}
\end{center}
\end{figure}
%%%%%%%%%%%%%%%%%%%%%%%%%%%%%%%%%%%%%%%%%%%%%%%%%%%%%%%%%%%%%%%%%%%

The holographic dark energy model has been strictly constrained by
the current cosmological observations. The joint analysis of the
latest observational data including the Constitution sample of 397
SN, the shift parameter of the CMB given by the five-year Wilkinson
Microwave Anisotropy Probe (WMAP5) observations, and the BAO
measurement from the Sloan Digital Sky Survey (SDSS), shows that the
parameter $c$ is indeed less than 1 (at nearly 2 $\sigma$ confidence
level):
$c=0.818^{+0.113}_{-0.097}~(1\sigma)~^{+0.196}_{-0.154}~(2\sigma)$
\cite{holoobs09}. Using the fitting result of Ref.~\cite{holoobs09},
we generate in Fig.~\ref{fig:wz} the evolution of the equation of
state of holographic dark energy $w(z)$ with 1$\sigma$ uncertainty.
In this figure, the central black solid line represents the best
fit, and the red dotted area around the best fit covers the range of
1$\sigma$ errors. The errors quoted in Fig.~\ref{fig:wz} are
calculated using a Monte Carlo method where random points are chosen
in the 1$\sigma$ region of the parameter space shown in Fig.~1 of
Ref.~\cite{holoobs09}. From Fig.~\ref{fig:wz}, one can clearly see
that our universe is striding into the phantom regime in the
holographic dark energy scenario. So, according to the current
observational data, it seems that the cosmic doomsday is inevitable
in this scenario. However, after detailedly investigating the
theoretical foundation of the holographic dark energy, one might ask
such a question: Is the cosmic doomsday really allowed in the
holographic scenario?

%\subsection{Is the cosmic doomsday allowed in the holographic dark
%energy model?}
\subsection{The physical motivation: Why must eliminate the big-rip singularity
in the holographic dark energy model?}\label{subsec:motivation}

First of all, it should be necessary to review the theoretical
foundation of the holographic dark energy model. In fact, the
holographic dark energy model is based on the effective quantum
field theory with some UV/IR relation \cite{Cohen:1998zx,Li:2004rb}.
Obviously, this is not the case in particle physics. In general, for
particle physics, one only needs an effective field theory with a UV
cutoff. It is usually assumed that the properties of elementary
particles can be accurately described by an effective field theory
with a UV cutoff less than the Planck mass $M_{pl}$, provided that
all momenta and field strengths are small compared with this cutoff
to the appropriate power. The standard model of particle physics
provides a good example for this. However, when gravity is
considered in the system, especially when black holes are involved,
the underlying theory of nature is suggested to be not a local
quantum field theory. Under the situation that a complete theory of
quantum gravity is not available, a good way of accurately
describing the world is to try to use the effective quantum field
theory in which the effects of gravity are adequately taken into
account and the range of validity for the local effective field
theory is determined. To accomplish this, some relationship between
UV and IR cutoffs should be imposed. Of course, the proposed IR
bound should not conflict with any current experimental success of
quantum field theory.

Local quantum field theory could not be a good effective low energy
description of any system containing a black hole, and should not
attempt to describe particle states whose volume is smaller than
their corresponding Schwarzschild radius. Nevertheless, in an
effective quantum field theory, for any UV cutoff $\Lambda$, there
is an sufficiently large volume for which the vastly overcounted
degrees of freedom of the effective field theory would lead to the
formation of a black hole spoiling the effective local quantum field
theory. To avoid this difficulty, Cohen et al. \cite{Cohen:1998zx}
propose a constraint on the IR cutoff $1/L$ which excludes all
states that lie within their Schwarzschild radius, namely,
$L^3\Lambda^4\lesssim L M_{pl}^2$. This bound implies a UV/IR
duality since the IR cutoff scales like $\Lambda^{-2}$. So, in fact,
they propose an effective local quantum field theory in which the UV
and IR cutoffs are not independent and only those states that can be
described by conventional quantum field theory are considered. The
holographic dark energy model is nothing but such an effective
quantum field theory with the IR cutoff length scale chosen to be
the size of the event horizon of the universe \cite{Li:2004rb}. In
this model, the UV cutoff $\Lambda$ runs with the cosmological
evolution since the event horizon of the universe as the IR cutoff
scale varies. Therefore, the holographic dark energy is actually a
dynamical vacuum energy \cite{holode2}.

Let us have a look at the late-time evolution of the holographic
cosmos. For the future event horizon, from Eq.~(\ref{Ode}), we have
$R_{eh}=c/(H\sqrt{\Omega_{de}})$. Hence, in the far future where the
dark energy totally dominates and other energy components are
diluted away, the event horizon behaves as $R_{eh}=cH^{-1}$. Using
Eqs.~(\ref{dotReh}) and (\ref{epsilon}), we can further obtain
$\dot{R}_{eh}=c-1$. Now, one can clearly see that the parameter $c$
plays a crucial role in determining the ultimate fate of the
universe: when $c=1$, we have $\dot{R}_{eh}=0$ that corresponds to a
de Sitter spacetime; when $c>1$, $\dot{R}_{eh}$ is a positive
constant indicating that $R_{eh}$ increases with a constant change
rate; when $c<1$, we see that $\dot{R}_{eh}$ is a negative constant
implying that $R_{eh}$ diminishes with a constant change rate. Thus,
$c<1$ will lead to a big-rip singularity at which $R_{eh}$ reaches
zero and other quantities such as the cosmological scale factor $a$,
the dark energy density $\rho_{de}$ and the Hubble expansion rate
$H$ approach infinity. However, the appearance of the big rip is
definitely beyond the scope of the effective quantum field theory
described above. Prior to the big rip, the UV cutoff $\Lambda$ will
first exceed the Planck mass $M_{pl}$, and then even the IR cutoff
scale $R_{eh}$ will become smaller than the Planck length $l_{pl}$.
Obviously, such a super-Planck phenomenon will break down the
theoretical foundation of the holographic dark energy model. So, it
should be confessed that the holographic dark energy model
undoubtedly has some congenital flaw in the phantom regime in some
sense. In view of the successes of the holographic dark energy model
in explaining the theoretical puzzles of dark energy
\cite{Li:2004rb,Li:2008zq} and fitting the observational data
\cite{obsholode1,obsholode2,holoobs09}, one should remedy this model
in the high energy regime to make the holographic theory of dark
energy more consistent and successful. The key for this is to find
out a working mechanism to eliminate the big-rip singularity in this
model.

\subsection{How to erase the big-rip singularity?}\label{subsec:how}

It is anticipated that some unknown high-energy physical effects,
especially those from the quantum gravity, might play an important
role shortly before the would-be big rip. Hence, conventionally, we
expect that when considering the effects of some unknown physics at
high energies, the big rip would be erased in the holographic dark
energy model. Now a question naturally arises asking what kind of
effects appears likely to be the right mechanism of amending the
behavior of holographic dark energy in high energy regime. In this
subsection, we shall discuss this issue.

It is the value of the parameter $c$ that determines whether the big
rip occurs: the appearance of the big rip is closely related to the
fact of $c<1$. So, to rescue the holographic dark energy model, one
may expect that at ultra high energies some quantum gravity effects
would lead to some correction to the parameter $c$, which makes the
parameter $c$ effectively change to be equal to or greater than one.
That is to say, we must impose some mechanism that leads to the high
energy corrections to $c$ to realize that $c_{eff}\geq 1$ at high
energies, where the effective parameter $c_{eff}$ is presumed to be
of the form
\begin{equation}
c_{eff}(t)=c+{\rm correction~from~high~energies}.\label{ceff}
\end{equation}
For the case of $c<1$, when the holographic dark energy is in the
low energy regime, it is obvious that $c_{eff}\rightarrow c$; when
the holographic phantom energy enters the high energy regime, the
quantum gravity is expected to begin to impact and consequently $c$
gets a corresponding correction like Eq.~(\ref{ceff}). Of course,
when $c\geq 1$, there is no high energy regime for dark energy, so
we always have $c_{eff}=c$ for these cases.

In order to erase the big rip, $c_{eff}\geq 1$ prior to the would-be
big rip must be satisfied. However, the most natural anticipation is
presumed to be $c_{eff}\rightarrow 1$ for which the steady state (de
Sitter) finale will emerge. Moreover, such a finale is expected to
be an attractor solution. Can such a dramatic mechanism really be
found out and performed in the holographic dark energy scenario? In
the next section, we shall accomplish this picture by employing the
extra-dimension mechanism (braneworld scenario).

\section{Holographic dark energy in braneworld
cosmology}\label{sec:brane}

In the present section, we shall discuss the scenario of holographic
dark energy in Randall-Sundrum (RS) braneworld. We will find that
the cosmic doomsday could be avoided successfully and the de Sitter
finale would emerge as an attractor.

\subsection{Why braneworld?}\label{subsec:why}

In the holographic dark energy model, the effects of gravity have
been adequately considered by using the holographic principle to
impose a UV/IR relationship in an effective local quantum field
theory, however, another important aspect of the spacetime, extra
dimensions, is absent. When the energy scale of dark energy is low
enough, the effects from extra dimensions are absolutely negligible;
however, when the phantom energy density becomes enormously high,
the extra-dimension effects are expected to play a significant role.
Accordingly, in order to make the holographic dark energy model more
complete, the extra dimensions should also be considered. In
addition, another reason for supporting the involvement of the
effects of extra dimensions in the holographic dark energy model
comes from the fact that the braneworld scenario might provide us
with some positive correction to $c$ as discussed in
Sec.~\ref{subsec:how}. So, it is quite interesting to study how the
physics of extra dimensions may affect the behavior of the
holographic dark energy in phantom regime. In the following we will
focus on a simple braneworld case.

Considering the case with one extra dimension compactified on a
circle, the effective four-dimensional Friedmann equation is
\cite{Randall:1999ee,Langlois:2002bb}
\begin{equation}
3H^2 = \rho \left(1 + \frac{\rho}{\rho_c} \right),\label{braneFeq}
\end{equation}
where $\rho_c=2\sigma$, with $\sigma$ the brane tension,
\begin{equation}
\sigma={6(8\pi)^2M_*^6\over M_{pl}^2},
\end{equation}
where $M_*$ is the true gravity scale of the five-dimensional
theory, and in this expression we explicitly write out $M_{pl}$. In
general, the most natural energy scale of the brane tension is of
the order of the Planck mass, but the problem can be generally
treated for any value of $\sigma>{\rm TeV}^4$.

It can be explicitly seen from the modified Friedmann equation
(\ref{braneFeq}) that the extra-dimension physics could contribute
some positive correction to the effective parameter $c_{eff}$,
namely,
\begin{equation}
c_{eff}(t)=c\sqrt{1+3c^2\rho_c^{-1} R_{eh}^{-2}(t)}.
\end{equation}
Therefore, for making the holographic dark energy model more
complete and consistent, it is quite natural to have recourse to the
extra dimension scenario.

\subsection{Cosmological evolution at late times}\label{subsec:braneeq}

In this subsection, we will discuss the late-time evolution of the
holographic dark energy in a braneworld, and derive the evolution
equation.

At the late times, the universe is totally dominated by the
holographic phantom energy, so in Eq.~(\ref{braneFeq}) we have
$\rho=\rho_{de}$. Note that for any cases
Eqs.~(\ref{dotReh})$-$(\ref{Odep}) are always satisfied since they
come from the definitions of holographic dark energy (\ref{density})
and event horizon (\ref{ehorizon}). In order to obtain the equation
of motion of dark energy, we need to calculate $\epsilon$ in
Eq.~(\ref{Odep}).

Taking the derivative of Eq.~(\ref{braneFeq}) with respect to time
$t$ and using the energy conservation equation
$\dot{\rho}_{de}=-3H(1+w)\rho_{de}$, we can get
\begin{equation}
\epsilon={3\over2}\Omega_{de}(1+w)\left(1+2{\rho_{de}\over\rho_c}\right),\label{eps2}
\end{equation}
where the fractional density of dark energy is still defined as
$\Omega_{de}=\rho_{de}/(3H^2)$, hence we have
\begin{equation}
{\rho_{de}\over\rho_c}={1\over\Omega_{de}}-1.\label{rrc}
\end{equation}
Substituting Eqs.~(\ref{eos}) and (\ref{rrc}) into
Eq.~({\ref{eps2}), we thus get
\begin{equation}
\epsilon=(2-\Omega_{de})\left(1-{1\over
c}\sqrt{\Omega_{de}}\right).\label{epsilonbrane}
\end{equation}

Substituting Eq.~(\ref{epsilonbrane}) into Eq.~(\ref{Odep}), we
eventually obtain the equation of motion for holographic phantom
dark energy ($c<1$) in a braneworld,
\begin{equation}
\Omega'_{de}=2\Omega_{de}(1-\Omega_{de})\left(1-{1\over
c}\sqrt{\Omega_{de}}\right).\label{eombrane}
\end{equation}
This differential equation governs the dynamical evolution of the
holographic dark energy when the effects of extra dimension begin to
play an important role. Therefore, the whole picture of the
holographic dark energy model is actually jointed by the two
segments: before the extra dimension effects work, the dynamics of
the holographic dark energy is dictated by Eq.~(\ref{eom}); after
the extra dimension effects emerge, the holographic dark energy
model will be governed by Eq.~(\ref{eombrane}). For the former
stage, the universe will finally be totally dominated by the phantom
energy, so at late times of this stage we have $\Omega_{de}=1$; this
is a stable attractor of Eq.~(\ref{eom}). When the evolution of the
phantom energy intrudes into the ultra high energy regime and the
extra dimension mechanism begins to operate, we find that
$\Omega_{de}$ begins to decrease with the expansion of the universe
though the phantom energy density $\rho_{de}$ always increases. From
Eq.~(\ref{rrc}), we see that assuredly $\Omega_{de}$ decreases with
the increase of $\rho_{de}$. Furthermore, from Eq.~(\ref{eombrane}),
one can find that $\Omega_{de}$ will decrease from 1 to a stable
value $c^2$ that corresponds to the final state of the universe, so
$\Omega_{de}=c^2$ should be the late-time attractor for the
holographic phantom energy in the RS braneworld.

\subsection{Finale of the universe: De Sitter spacetime
attractor}\label{subsec:ds}

We will show that the finale of the universe in this scenario is a
de Sitter (steady state) spacetime.

First, let us prove that $\Omega_{de}=c^2$ is a stable late-time
attractor solution to Eq.~(\ref{eombrane}). Considering a
perturbation to this solution, we find that the soultion will be
recovered soon: when $\Omega_{de}<c^2$, owing to that
$1-\Omega_{de}>0$ and $1-{1\over c}\sqrt{\Omega_{de}}>0$, we have
$\Omega'_{de}>0$ indicating that $\Omega_{de}$ will increase until
it reaches $c^2$; when $c^2<\Omega_{de}<1$, since $1-\Omega_{de}>0$
and $1-{1\over c}\sqrt{\Omega_{de}}<0$, we have $\Omega'_{de}<0$
implying that $\Omega_{de}$ will keep on decreasing until it touches
$c^2$. So, out of question, $\Omega_{de}=c^2$ is a stable late-time
attractor solution.

In this stage, the Hubble expansion rate will increase until it
becomes a constant. For convenience, we define
\begin{equation}
\tilde{h}^2\equiv {H^2\over \rho_c}={1-\Omega_{de}\over
3\Omega_{de}^2},
\end{equation}
and we find that the maximum of $\tilde{h}$ is
\begin{equation}
\tilde{h}_{max}={\sqrt{1-c^2}\over\sqrt{3}c^2},
\end{equation}
which corresponds to the late-time attractor solution
$\Omega_{de}=c^2$. Therefore, we can draw the conclusion that the
finale of the universe in this scenario is a de Sitter spacetime. At
the finale, using Eqs.~(\ref{ceff}) and (\ref{rrc}), one can check
that $c_{eff}=1$. Also, from Eq.~(\ref{ceff}), one derives the
minimum of the size of the event horizon,
\begin{equation}
R_{eh}^{min}={\sqrt{3}c^2\over\sqrt{(1-c^2)\rho_c}}.\label{ehmin}
\end{equation}
By far, in this modified holographic dark energy model (with $c<1$),
it is of interest to find that the universe begins with an inflation
and also ends with another inflation.

%%%%%%%%%%%%%%%%%%%%%%%%%%%%%%%%%%%%%%%%%%%%%%%%%%%%%%%%%%%%%%%%%%
\begin{figure}[htbp]
\begin{center}
\includegraphics[scale=0.3]{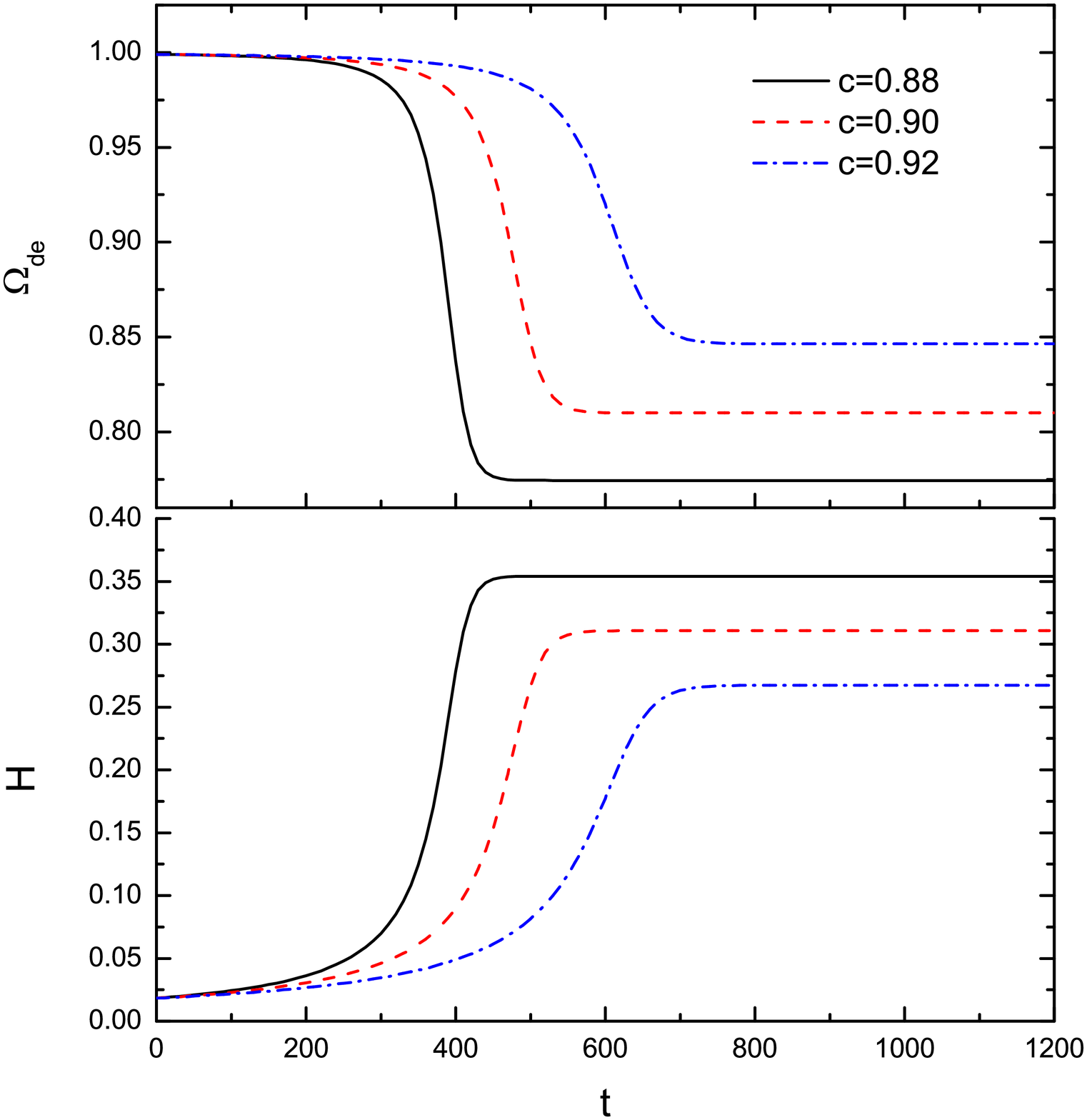}
\caption{The late-time evolution of $\Omega_{de}$ and $H$ in the
scenario of holographic dark energy in a RS braneworld. Here we take
three cases as example, namely, $c=0.88$, 0.90 and 0.92. From this
figure, we can explicitly see that the finale of the universe in
this scenario is a steady-state spacetime. Note that in this plot
the Hubble expansion rate $H$ is in units of
$(\sqrt{\rho_c}/M_{pl})$ and the cosmic time $t$ is in units of
$(M_{pl}/\sqrt{\rho_c})$. }\label{fig:final}
\end{center}
\end{figure}
%%%%%%%%%%%%%%%%%%%%%%%%%%%%%%%%%%%%%%%%%%%%%%%%%%%%%%%%%%%%%%%%%%%

As an example, we plot the late-time evolution curves of
$\Omega_{de}$ and $H$ according to the solution of the differential
equation (\ref{eombrane}), as shown in Fig.~\ref{fig:final}. The
initial condition of the calculation of Eq.~(\ref{eombrane}) is
taken as $\Omega_{de}=1^-$ when $t=0$. Here we take three cases as
example, namely, $c=0.88$, 0.90 and 0.92. From this figure, we can
explicitly see that the finale of the universe in this scenario is a
steady-state spacetime. Note that in this plot the Hubble expansion
rate $H$ is in units of $(\sqrt{\rho_c}/M_{pl})$ and the cosmic time
$t$ is in units of $(M_{pl}/\sqrt{\rho_c})$.

So far, we have constructed a complete model of holographic dark
energy in which the evolution of the universe is divided into two
stages: In the low-energy regime, the dynamical evolution of the
universe is governed by Eq.~(\ref{eom}); in the high-energy regime,
the dynamical evolution of the universe is dictated by
Eq.~(\ref{eombrane}). Therefore, the ultimate fate of the universe
in the case of $c<1$ should be a steady-state spacetime, in stead of
a cosmic doomsday. The consideration of an extra-dimension mechanism
provides a successful solution to the theoretical puzzle of the
holographic dark energy model in the presence of a cosmic doomsday.
Moreover, in quantum gravity theories such as the string/M theory,
the spacetime is commonly believed to be fundamentally higher
dimensional, so the involvement of the extra dimensions makes the
holographic dark energy model more complete. In addition, it should
be noted that the high-energy regime only exists in the case of
$c<1$, and in the cases of $c\geq 1$ there is no high-energy regime
so that the extra-dimension effects will be absent in those cases.

It is conceivable that the parameter $c$ of the holographic dark
energy may be constrained by the braneworld cosmology. The IR cutoff
length scale $R_{eh}$ in the effective quantum field theory
describing the holographic dark energy gets a minimum when the
universe goes into the steady-state finale, as given by
Eq.~(\ref{ehmin}). If we impose a plausible additional requirement
on $R_{eh}^{min}$ that the IR cutoff length scale should be greater
than the compactification radius of the extra dimensional circle,
namely, $R_{eh}^{min}\gtrsim l$, where $l=2\sqrt{3}\rho_c^{-1/2}$ is
the anti de Sitter length scale, we will derive the constraint
$c\gtrsim 2(\sqrt{2}-1)$ (namely, $c\gtrsim 0.91$) that is
consistent with the fitting result of the observational data
\cite{holoobs09}. It should also be noted that this constraint
result should not be taken so seriously because actually the
additional requirement on IR cutoff does not seem to be necessary.

One may also consider anther possible braneworld scenario in which
the effective four-dimensional Friedmann equation is
$3H^2=\rho(1-\rho/\rho_c)$, where the negative sign arises from a
second timelike dimension \cite{Shtanov:2002mb}. Note that this
modified Friedmann equation can also arise from the loop quantum
cosmology. Such a modified Friedmann equation with matter and
phantom energy components can lead to a cyclic universe scenario in
which the universe oscillates through a series of expansions and
contractions \cite{Brown:2004cs}. Can this braneworld scenario help
to eliminate the big-rip singularity in the holographic dark energy
model? In fact, the high energy correction to $c$ in this case is
negative, namely, $c_{eff}=c\sqrt{1-3c^2\rho_c^{-1}R_{eh}^{-2}}$, so
the cosmic doomsday finale could not be replaced by a steady state
one in this case. One may further conceive that perhaps the
holographic dark energy model combined with this scenario would
replace the big-rip singularity with a turnaround. However, this is
also impossible since a cyclic universe has no a future event
horizon such that the definition of holographic dark energy breaks
down in this scenario. In Ref.~\cite{holocycl}, such a scenario is
investigated, but to evade this difficulty the future event horizon
is redefined. So, it should be confessed that the scenario discussed
in Ref.~\cite{holocycl} is not realistic but only a toy model. The
steady state future of the holographic Ricci dark energy in RS
braneworld has been discussed in Ref.~\cite{RicciRS}. Other
scenarios describing the holographic dark energy in braneworld can
be found in, e.g., Refs.~\cite{Wu:2007tp}.

In addition, to avoid the big rip, it also seems quite natural to
consider some interaction between holographic dark energy and matter
\cite{Li:2008zq,intholo}. However, from the definition of the
holographic dark energy (\ref{density}), one can see that it is
closely related to the future event horizon $R_{eh}$ that is a
global concept of spacetime. So, unlike the interaction between
scalar-field dark energy and dark matter \cite{intde}, it is rather
difficult to imagine the local interaction between holographic dark
energy and matter. Therefore, we feel that the extra dimension
recipe is better than the interaction one for solving the big rip
crisis in the holographic dark energy model.

\subsection{The whole story: From past to future}

Finally, let us derive the evolution equation describing the whole
expansion history from the past to the far future for the
holographic dark energy model in a RS braneworld.

Consider, now, a general case that the universe contains dust matter
(dark matter plus baryons) and holographic dark energy, namely,
$\rho=\rho_m+\rho_{de}$, where $\rho_m=\rho_{m0}a^{-3}$ and
$\rho_{de}=\rho_{de0}f(a)$ with $f(a)=\rho_{de}(a)/\rho_{de0}$. Note
that here the subscript ``0'' marks the quantities corresponding to
today, and for the scale factor of the universe $a$ we have let
$a_0=1$. From the modified Friedmann equation (\ref{braneFeq}), one
can easily derive
\begin{equation}
E(a)^2=\left(\Omega_{m0}a^{-3}+\Omega_{de0}f(a)\right)
\left[1+\beta\left(\Omega_{m0}a^{-3}+\Omega_{de0}f(a)\right)\right],\label{Ea}
\end{equation}
where $E\equiv H/H_0$, $\Omega_{m0}=\rho_{m0}/(3H_0^2)$,
$\Omega_{de0}=\rho_{de0}/(3H_0^2)$, and $\beta=3H_0^2/\rho_c$. The
dimensionless parameter $\beta$ characterizes the ratio of the
present-day density $\rho_0$ to the critical density of the
braneworld $\rho_c$. From the fact that $\rho_0\simeq 4\times
10^{-47}~{\rm GeV}^4$ and $\rho_c$ takes some value between
$10^{12}$ and $10^{72}$ ${\rm GeV}^4$, we estimate that the value of
$\beta$ lies between $10^{-119}$ and $10^{-59}$. For a given
$\Omega_{m0}$, one can get
$\Omega_{de0}=(\sqrt{1+4\beta}-1)/2\beta-\Omega_{m0}$ from
Eq.~(\ref{Ea}) by using the conditions $E_0=E(t_0)=1$ and
$f_0=f(t_0)=1$. Also, by definition, the fractional densities of
matter and dark energy can be expressed as
\begin{equation}
\Omega_m={\rho_m\over 3 H^2}={\Omega_{m0}a^{-3}\over
E^2},~~~\Omega_{de}={\rho_{de}\over 3 H^2}={\Omega_{de0} f(a) \over
E^2}.\label{OmOde}
\end{equation}

%%%%%%%%%%%%%%%%%%%%%%%%%%%%%%%%%%%%%%%%%%%%%%%%%%%%%%%%%%%%%%%%%%
\begin{figure}[htbp]
\begin{center}
\includegraphics[scale=0.35]{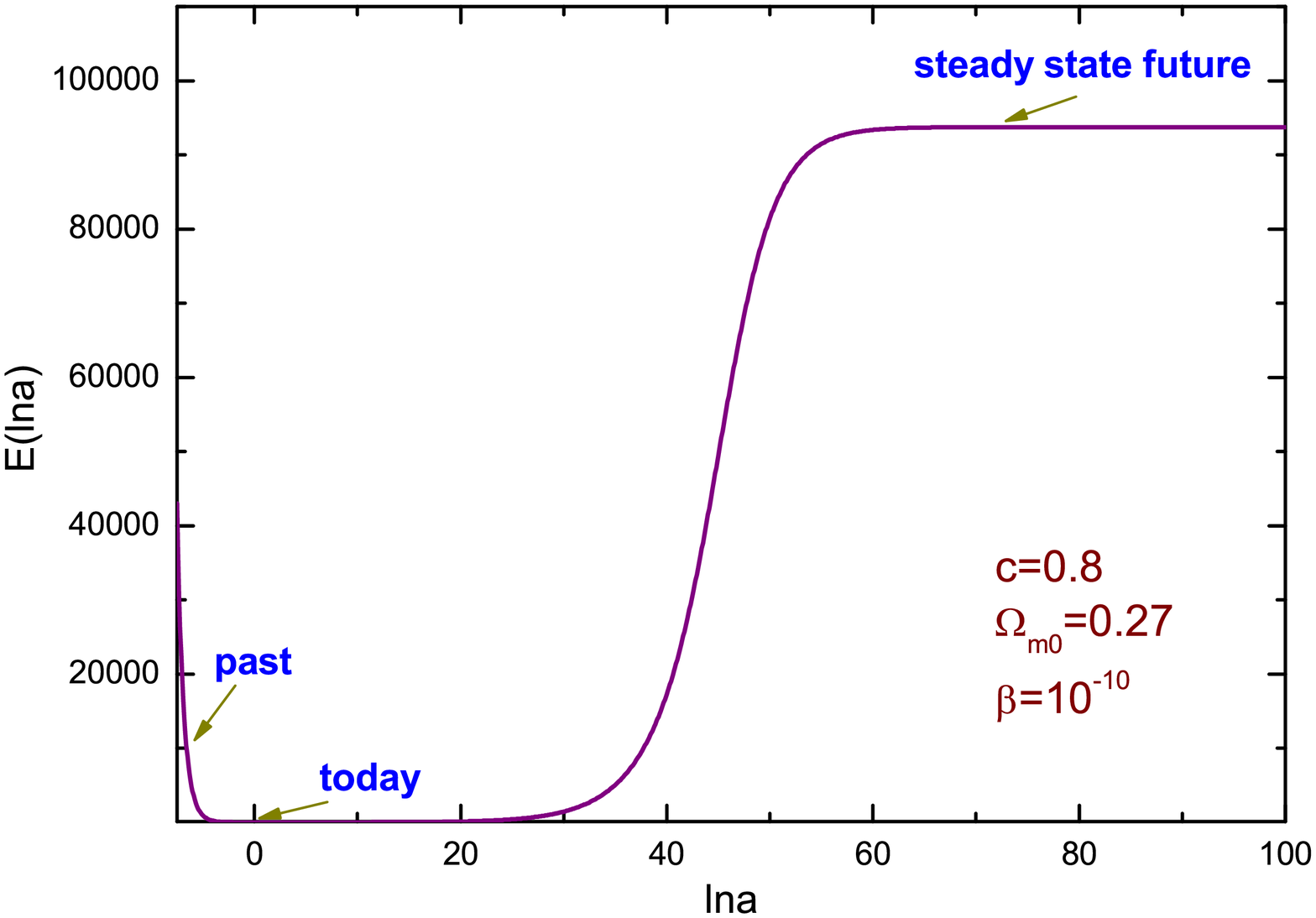}
\caption{The expansion history from the past to the far future for
the holographic dark energy model in a RS braneworld. As a schematic
example, here we take the case $c=0.8$, $\Omega_{m0}=0.27$ and
$\beta=10^{-10}$. Note that in this example the values of $c$ and
$\Omega_{m0}$ are realistic but the value of $\beta$ is evidently
unrealistic because it is much bigger than a reasonable value that
is in the range of $10^{-119}-10^{-59}$. This is only for the
display effect of the plot.}\label{fig:whole}
\end{center}
\end{figure}
%%%%%%%%%%%%%%%%%%%%%%%%%%%%%%%%%%%%%%%%%%%%%%%%%%%%%%%%%%%%%%%%%%%

From Eq.~(\ref{braneFeq}) one can calculate $\epsilon=-\dot{H}/H^2$
and obtains
\begin{equation}
\epsilon={3\over
2}[(1+w)\Omega_{de}+\Omega_m]\left(1+2{\rho\over\rho_c}\right),
\end{equation}
and also from Eq.~(\ref{braneFeq}) one derives
\begin{equation}
{\rho\over\rho_c}={1\over\Omega_{de}+\Omega_{m}}-1.
\end{equation}
Consequently, in combination with Eq.~(\ref{eos}) one obtains
\begin{equation}
\epsilon=\left({3\over 2}\Omega_m+\Omega_{de}-{1\over
c}\Omega_{de}^{3/2}\right)\left({2\over
\Omega_{de}+\Omega_m}-1\right).\label{epsilon3}
\end{equation}
To get the evolutionary behavior of holographic dark energy $f(a)$,
one only needs to substitute Eq.~(\ref{epsilon3}) into
Eq.~({\ref{Odep}) and solve the derived differential equation with
the initial condition $f(t_0)=1$. So far, we can describe the whole
story of the holographic dark energy in a RS braneworld by using
Eqs.~(\ref{Odep}), (\ref{Ea}), (\ref{OmOde}) and (\ref{epsilon3}).

As a schematic example, we take the case $c=0.8$, $\Omega_{m0}=0.27$
and $\beta=10^{-10}$, and plot the expansion history of the
universe, namely, $E(\ln a)$, in Fig.~\ref{fig:whole}. Note that in
this example the values of $c$ and $\Omega_{m0}$ are realistic but
the value of $\beta$ is evidently unrealistic because it is much
bigger than a reasonable value that is in the range of
$10^{-119}-10^{-59}$. This is only for the display effect of the
plot. We remind the reader to notice the evolutionary trend of the
universe shown in Fig.~\ref{fig:whole} but forget the unreality as a
schematic example. The steady state (de Sitter) future can be
explicitly identified in this figure. Also, it is shown in this
figure that the past and the future of the expansion history in this
scenario can be seamless linked.

\section{Conclusion}\label{sec:conclusion}

The holographic principle plays a very significant role in studying
the dark energy problem. With the consideration of the holographic
principle, the holographic dark energy model is proposed by
constructing an effective quantum field theory with a UV/IR duality.
Thanks to the UV/IR correspondence, the UV problem of dark energy
can be converted into an IR problem. In the holographic dark energy
model, the IR cutoff length scale is chosen as the size of the event
horizon of the universe.

According to the observational data, it seems inevitable that the
cosmic doomsday would be the ultimate fate of the universe in the
holographic dark energy model. However, unfortunately, the big-rip
singularity will undoubtedly ruin the theoretical foundation of the
holographic dark energy scenario that is based upon an effective
quantum field theory. To rescue the holographic dark energy model,
we employ the braneworld cosmology and incorporate the
extra-dimension effects into this model. The motivation of
considering the extra dimensions consists of two aspects: (a) The
spacetime is commonly believed to be fundamentally higher
dimensional in quantum gravity theories such as string/M theory, so
the involvement of the extra dimensions would make the holographic
dark energy model more complete. (b) With the help of the extra
dimension mechanism, the big-rip singularity in the holographic dark
energy model could be erased successfully.

In this paper, we have investigated the cosmological evolution of
the holographic dark energy in the braneworld cosmology. It is of
interest to find that for the far future evolution of the
holographic dark energy in a RS braneworld, there is a late-time
attractor solution where the steady state (de Sitter) finale occurs,
in stead of the big rip. Therefore, in the holographic dark energy
model, the extra-dimension recipe could heal the world.

\begin{acknowledgments}
The author is grateful to Qing-Guo Huang, Miao Li, Yun-Song Piao and
Yi Wang for helpful discussions. This work was supported by the
Natural Science Foundation of China under Grants Nos.~10705041 and
10975032.
\end{acknowledgments}

\end{document}